\documentclass[aip,jmp,amsmath,amssymb,floatfix]{revtex4-1}

\usepackage{graphicx}
\usepackage{dcolumn}
\usepackage{bm}
\usepackage[version=3]{mhchem}

\begin{document}


\title[]{Simple Locking of IR and UV diode lasers to a visible laser using a LabVIEW PID controller on a Fabry-Perot signal}

\author{J.M. Kwolek}
	
\author{J.E. Wells}
\affiliation{Physics Department, University of Connecticut, Storrs, Connecticut 06269}
\author{D.S. Goodman}
\affiliation{Physics Department, University of Connecticut, Storrs, Connecticut 06269}
\affiliation{Physics Department, Wentworth Institute of Technology, Boston, Massachusetts 02115}
\author{W.W. Smith}
\email{Winthrop.Smith@uconn.edu}
\affiliation{Physics Department, University of Connecticut, Storrs, Connecticut 06269}

\date{\today}

\begin{abstract}
Simultaneous laser locking of infrared and ultraviolet lasers to a visible stabilized reference laser is demonstrated via a Fabry-Perot (FP) cavity. LabVIEW is used to analyze the input and an internal proportional-integral-derivative algorithm converts the FP signal to an analog locking feedback signal. The locking program stabilized both lasers to a long term stability of better than 9 MHz, with a custom-built IR laser undergoing significant improvement in frequency stabilization. The results of this study demonstrate the viability of a simple, computer-controlled, non temperature-stabilized FP locking scheme for our application, laser cooling of $\ce{Ca+}$ ions, and its use in other applications with similar modest frequency stabilization requirements.
\end{abstract}

\maketitle

\section{Introduction}

Laser locking is a fundamental method in atomic physics research, but achieving the required precision can be complex. Lasers used to cool and trap neutral atoms are typically locked to atomic lines using Doppler-free saturation spectroscopy. Commonly used atomic and molecular species in this type of apparatus include the alkalis, iodine, and molecular tellurium-130\cite{tellurium}. With the exception of potassium at 404 nm, these species lack lines around 400 nm. For example, the cooling transition for $\ce{Ca+}$ ions\cite{caion} is at 397 nm, which is too large a wavelength difference from the neutral potassium line for an offset lock to the potassium saturation spectrum to be a viable locking technique.

There are many techniques of varying difficulties to lock lasers, such as optical feedback locking\cite{Hayasaka:2002}, Doppler-free spectroscopy\cite{Park:2003, revise3}, locking to a stabilized reference laser in a transfer cavity\cite{toyoda, Reich:1986, Nicolas:1989, Xie:1989,Riedle:1993,transfer,transfer2}, and locking to ion transitions via a hollow cathode lamp\cite{revise1, revise2}. These methods, while effective, each present their own unique challenges to implement. All of these platforms require a feedback system to lock the laser, while those that use temperature-stabilized Fabry-Perot (FP) cavities or transfer cavities require additional feedback systems for the cavities themselves. These extra feedback systems add to the complexity of these locking methods.

We present a study of a simple method of locking a laser to a highly stable reference laser within a single FP cavity and a single software proportional-integral-derivative (PID) regulator. Many contemporary atomic physics experiments require multiple lasers, and it is common for atomic physics laboratories to use at least one stabilized laser system. The method presented here enables us to lock both ultraviolet (UV) and infrared (IR) laser diodes to our reference laser using widely available dichroic mirrors, FP cavities, and National Instruments' (NI) hardware and software (LabVIEW). We take advantage of our stabilized laser at 589 nm used for the cooling and trapping of neutral sodium atoms (which is already stabilized to an atomic standard) to lock the ion-cooling laser at 397 nm and the repumper laser at 866 nm.   

The locking method presented here is similar to those that lock a FP cavity to a stabilized reference beam (transfer cavity)\cite{Reich:1986, Nicolas:1989, Xie:1989,Riedle:1993,transfer,transfer2}. In the transfer cavity lock, the cavity is locked to an independantly stabilized reference laser first, then the secondary laser is locked to the cavity. To achieve independant stabilization of the Fabry-Perot (FP) cavity requires additional equipment, usually including an electro-optic modulator and independant PID circuit. P.~Bohlouli-Zanjani et~al. created a lock using a transfer cavity with a stability of $\sim 1 $MHz.\cite{transfer} We achieve comparable stability in a simpler way. With a more sophisticated setup, F.~Rhode et~al. obtained a lock stability of 130 kHz by using custom-built analog PID modules in addition to the required modulators.\cite{transfer2} However, many laser locking applications do not require such a robust lock, such as the Doppler cooling of ions (which often have wider atomic transition linewidths as compared to neutral alkali atoms). Thus, we propose a straightforward digital lock where the FP cavity length is continuously scanned and the two laser signals are compared with each other. This simplifies the lock setup significantly and eliminates the need for any temperature stability or extra feedback systems.

Nizamani, et~al.\cite{Nizamani:2013} also use a continuously scanned FP cavity, but in that case, two IR wavelengths are locked to one another. The beams have different polarizations and are combined using polarizing beamsplitters (PBS).  However, optics designed for one wavelength regime are often ineffective at other wavelengths. For this reason, our method uses dichroic mirrors. An additional notable outcome of our work is that it shows how easily this technique can be to generalized to lock very different wavelengths or even multiple lasers to the same reference laser.

For most uses, locking UV lasers with a hollow cathode lamp is a sufficient atomic reference to frequency lock, as in the experiments with E. W. Streed et~al.\cite{revise1} and S. C. Burd et~al.\cite{revise2}. In addition, J. Smith et~al\cite{revise3} were able to lock using an atomic vapor cell reference. However, these schemes are limited by the Stark and Doppler-broadened linewidth width of the transition, which is sometimes too broad for locking. Also, locking with a hollow cathode lamp is very specific, only allowing locking to specific ionic transitions. Our use of an offset lock allows for locking to any wavelength, simply limited by the optics in the FP cavity. 

Our paper is organized as follows: First we discuss the experimental setup, including the optical and data acquisition (DAQ) components of our system. Second, we provide an overview of the LabVIEW interface used and the programming element of our system. Third, we characterize the resulting lock stability to determine the viability of this method for our stability requirements.

\section{Experimental Setup}

\subsection{Equipment}

The diode lasers used by our research group are a narrow-linewidth Toptica DL-SHG pro, a Toptica DL100, and a custom-built external cavity diode laser, operating at 589 nm, 397 nm, and 866 nm respectively. Our group uses the 589 nm laser for trapping and cooling of sodium, which is frequency stabilized with an atomic Na vapor saturated absorption setup.\cite{{smith1},{smith2},{smith3}} The long-term frequency stability of the 589 nm laser is narrow (150 kHz), as compared to any of the Na D2 transition linewidths. We use this laser as a reference beam for the locking the other lasers. We lock the 397 nm and 866 nm laser with separate Thorlabs SA200 FP interferometers (free spectral range of 1.5 GHz) and four photodiodes from Advanced Photonix Inc. (SD100-13-23-222-ND, (2) SD100-12-22-021-ND, SD100-14-21-021-ND), two for each wavelength pair.

\begin{figure}
	\vspace{1em}
	\includegraphics[width=0.9\linewidth]{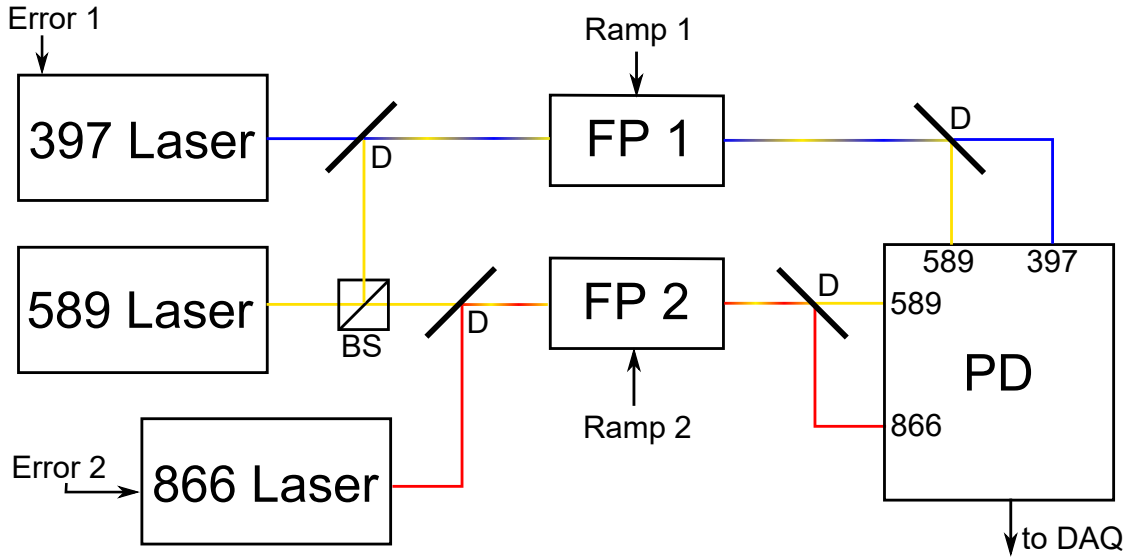}
	\caption{\label{fig:table} (Color Online) The optical table contains three lasers, two of which are paired with the 589 nm reference laser. The frequency stabilization of the 589 nm laser itself is not shown here. Each pair of beams is combined with a dichroic mirror (D) before being sent through its respective FP cavity (FP 1,2) and then separated by a second dichroic mirror. The four photodiodes (PD) convert the light into a current signal which is then sent to the DAQ system and a control computer via current-to-voltage conversion in the PD box. Each arrow represents an electronic connection, and BS represents a visible beamsplitter.
	}
\end{figure}

\begin{figure}
	\vspace{1em}
	\includegraphics[width=0.9\linewidth]{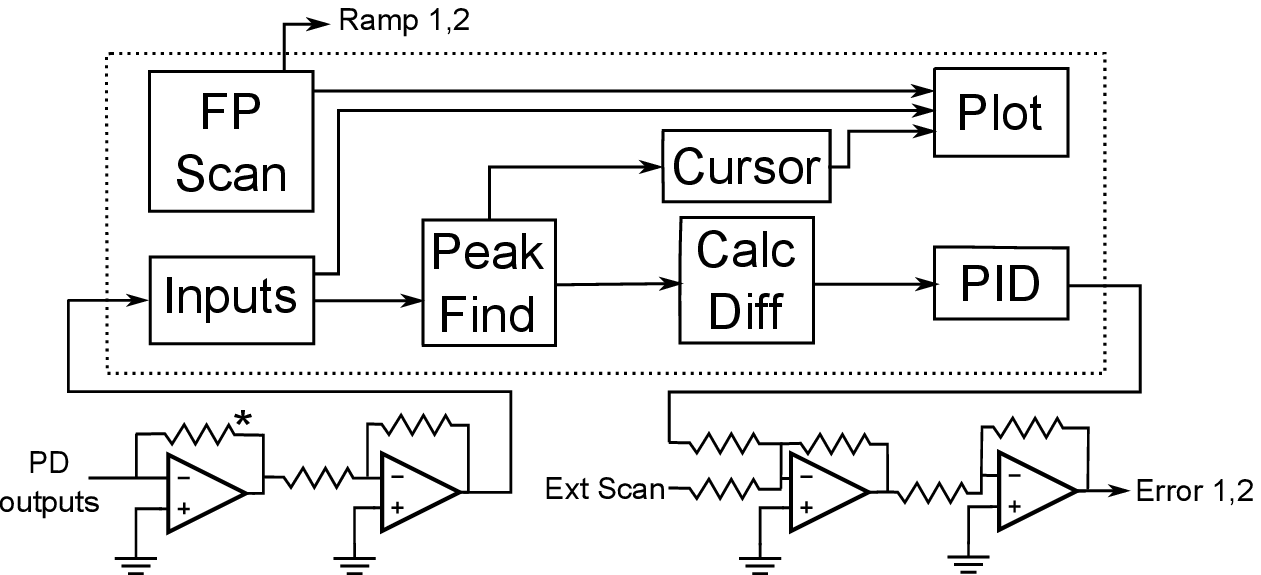}
	\caption{\label{fig:labview}The LabVIEW program reads the voltage measurements from the photodiodes (PD), analyzes the peaks, and outputs an analog error signal signal which is added to any additional externally controlled offsets. The total error signal is then fed to the diode lasers piezo-controlled diffraction grating. The starred resistor controls the input gain. All other resistors have the same equal values.
	}
\end{figure}

We create two sets of colinear beams: one 589-866 nm beam and one 589-397 nm beam, which are directed into separate FP cavities, labeled FP1 and FP2 in Fig.~\ref{fig:table}. After the light passes through the cavities, a dichroic mirror separates the two wavelengths and the transmitted intensity of each wavelength is separately measured by wavelength specific photodiodes. An operational-amplifier (op-amp) current-to-voltage circuit is used to change each photodiode current measurement into a voltage before feeding the signal into our National Instruments (NI) DAQ hardware, as seen in Fig.~\ref{fig:labview}. Using the circuit to change the current measurement to a voltage measurement gives better impedance matching between the photodiode and the DAQ input, resulting in a cleaner input signal.

\begin{figure}
	\vspace{1em}
	\includegraphics[width=0.9\linewidth]{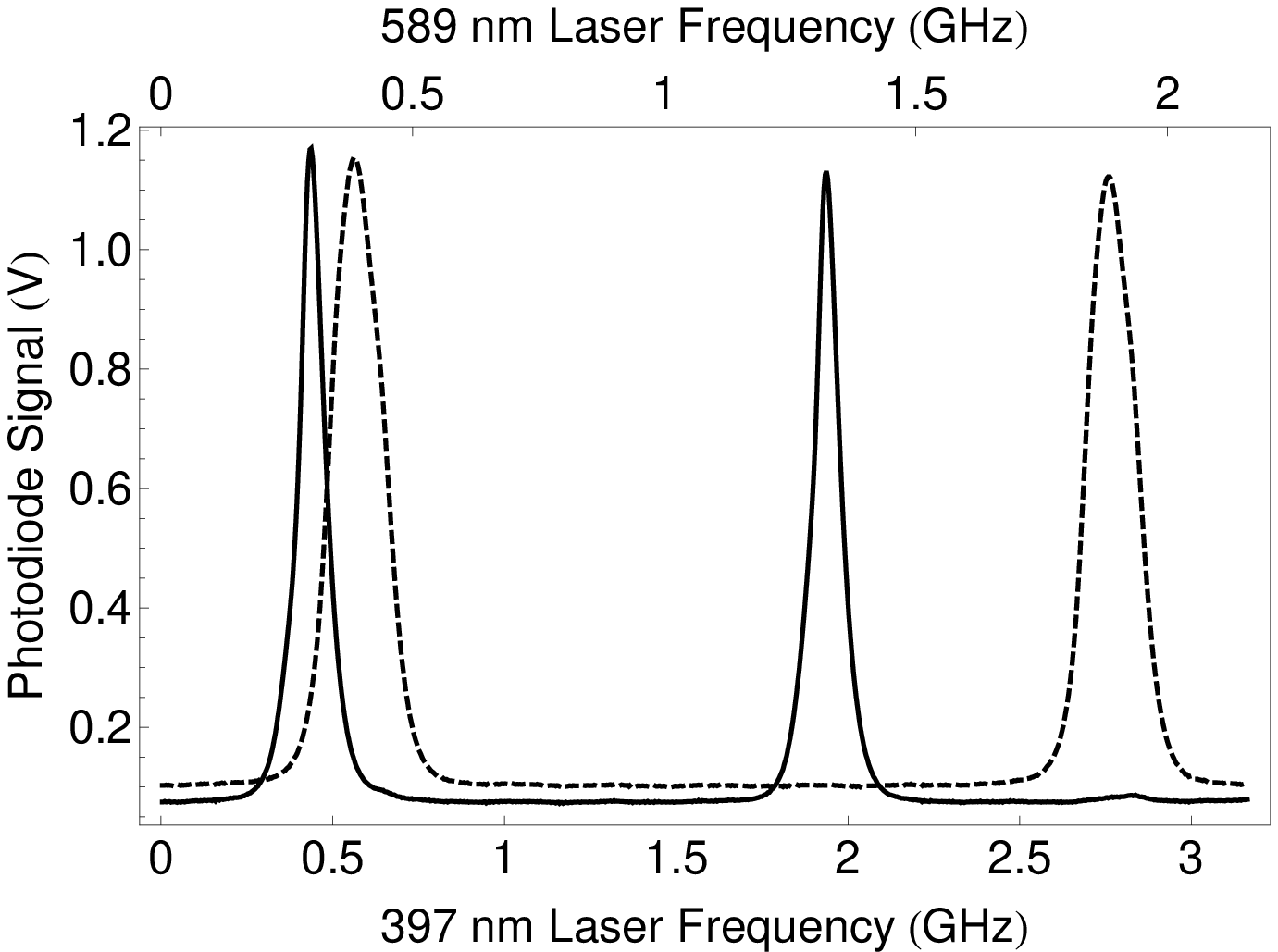}
	\caption{\label{fig:fpscan} Each FP ramp scans over resonances of both the lasers in the cavity. The solid signal is from the transmission of 397 nm laser, while the dashed signal is from the transmission of the 589 nm laser. The x-axes are expressed in terms of the relative laser frequency of each color. While scanning the cavity length, the peak interval varies in proportion to the laser wavelength, so the 589 nm peaks are farther apart than the 397 nm peaks. The linewidths of the 397 nm and 589 nm signal are 80.7(6) MHz and 101(2) MHz respectively.
	}
\end{figure}

As the length of each cavity is scanned with a sawtooth waveform, each FP cavity will transmit light when the cavity length is an integer multiple of the wavelength. We observe peaks from each of the lasers in the cavity, as shown in Fig.~\ref{fig:fpscan}. We obtain the peaks as a function of sample number, recorded by the LabVIEW VI, which varies directly with cavity length during a scan. We then plot as a function of the relative frequency of each laser, calibrating our horizontal scales to the free spectral ranges for each color as seen in Fig. \ref{fig:fpscan}.

The signal fed back to the diode lasers' piezo-controlled grating is the sum of the error signal generated via the locking algorithm (discussed in Sec. IIB) and an externally controlled offset. The addition of the signals is performed using an op-amp voltage adding circuit shown in Fig.~\ref{fig:labview}, which allows us to scan the diode laser frequencies while the laser is locked or add an external offset.

\subsection{LabVIEW Interface}

The LabVIEW interface incorporates an NI 9215 analog input board along with an NI 9205 output board in order to interact with the electronic elements on the optical table. The inputs and outputs are noted in Fig.~\ref{fig:table} as arrows entering or exiting each of the optical components. The program runs in a loop which is repeatedly executed until a stop command is given. It begins by initializing a continuously running scan of the FP cavities. The LabVIEW built-in virtual instrument (VI) peak-find algorithm determines the relative location of each laser's FP signal peak (in units of sample number) from its respective photodiode signal. The feedback loop tries to keep the relative transmission peak location constant. The continuously updated value of the relative peak locations is considered the input process variable for the LabVIEW built-in PID VI.

The PID VI is a digitally programmed and implemented PID regulator that works similarly to its analog counterpart with a set point, process variable, and error output\cite{LabVIEW}. In our case, the set-point is the desired peak separation and the process variable is the actual peak separation. The process variable is compared to the set-point and their difference is converted into an output via the typical PID formula adopted to our discrete system,
\begin{equation}
F(t)=C_p f(t)+C_i\sum_{\tau=0}^t f(\tau)+C_d\frac{\Delta f(t)}{\Delta t}.
\end{equation}
In this formula, $F(t)$ is the output value, $f(t)$ is the input process variable, and $C_k$ are the PID parameters. The quantity $\Delta t$ is set by the time of one iteration of the program.

\begin{figure}
	\includegraphics[width=0.9\linewidth]{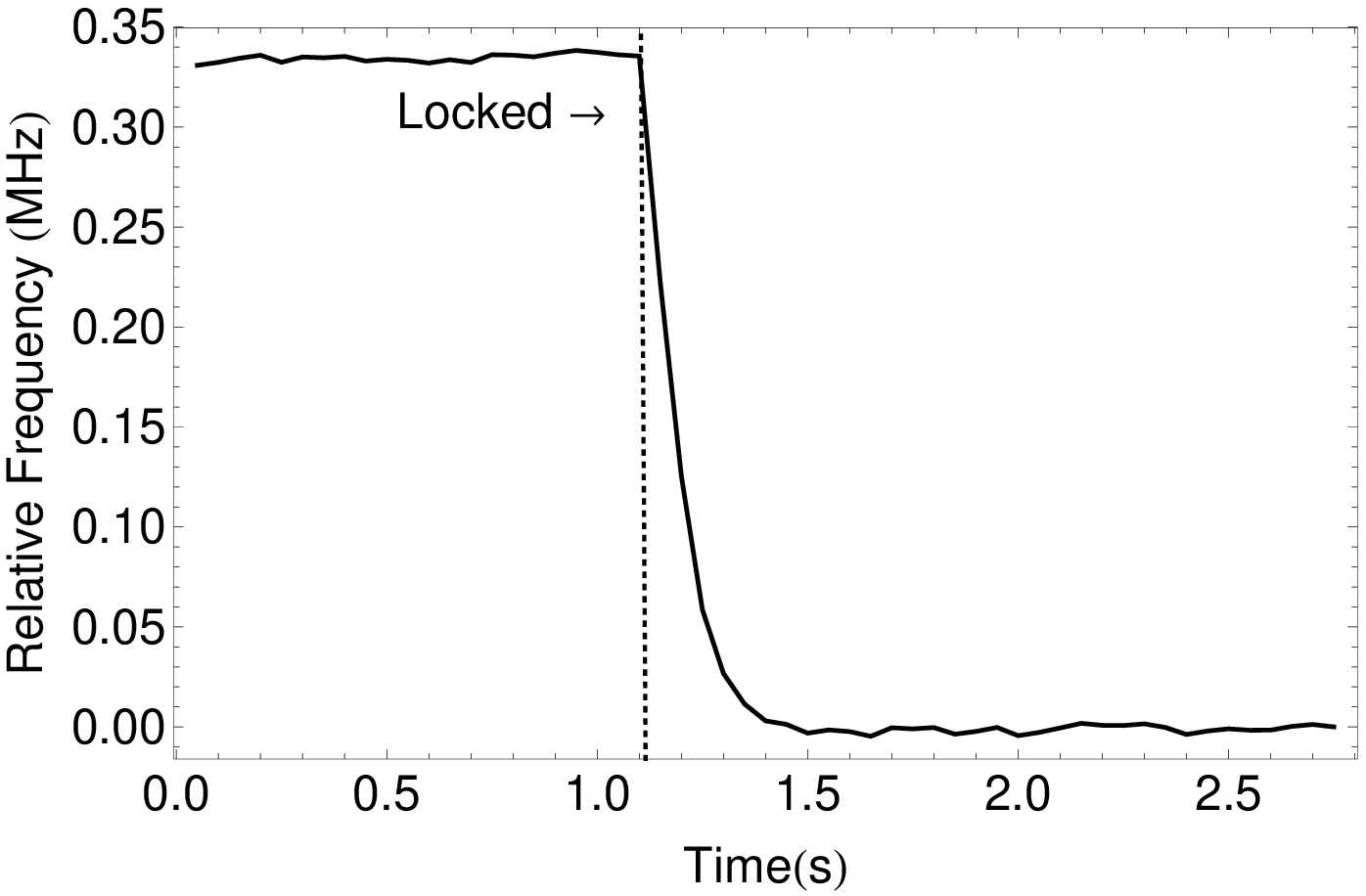}
	\caption{\label{fig:locking} The initial locking behavior of the 397 nm laser is demonstrated. The PID set point is set to a relative frequency difference of 0. In this figure, the laser locks to a steady state in less than 0.5 seconds.
	}
\end{figure}

For example, once the 397 nm laser is at the desired frequency and the lock is engaged, an error signal is generated that attempts to maintain the relative peak signal separation between the 589 nm FP signal peak and the 397 nm FP signal peak. The immediate locking behavior is demonstrated in Fig. \ref{fig:locking}. The relative position between the independently stabilized 589 nm signal and the 397 nm FP signal passing through a single FP cavity stays constant, even when the FP signals undergo thermal drift, because the two peaks drift together, so the relative position is a thermal-drift-insensitive lock point. This scheme requires that the length of the FP cavities be continuously scanned, so that the current FP signal peak difference can be continually measured and the error signal be continually updated once each iteration of the program's loop. Each iteration of the program takes roughly 160 ms, which is the limit to the speed of the lock. This entire locking process is pictured schematically in Fig.~\ref{fig:labview}. 

Our particular National Instruments (NI) DAQ analog input hardware runs at 100 kS/s (kilo-samples per second) per channel, which in practice limits our ability to resolve signals. However, NI boards with much higher rates, 1 MHz, do exist. This means, in our experiment, that the factor limiting our locking speed is the FP ramp and the speed of the LabVIEW program. Thorlabs sells separately a control box (that we do not use in our setup) to ramp their FP, which has a maximum scan rate of 100 Hz. In practice, it has been our experience that we can scan the FP faster than 100 Hz without damage to the FP. In principle, a more streamlined program could improve the speed of the lock. 

\begin{figure}
	\includegraphics[width=0.9\linewidth]{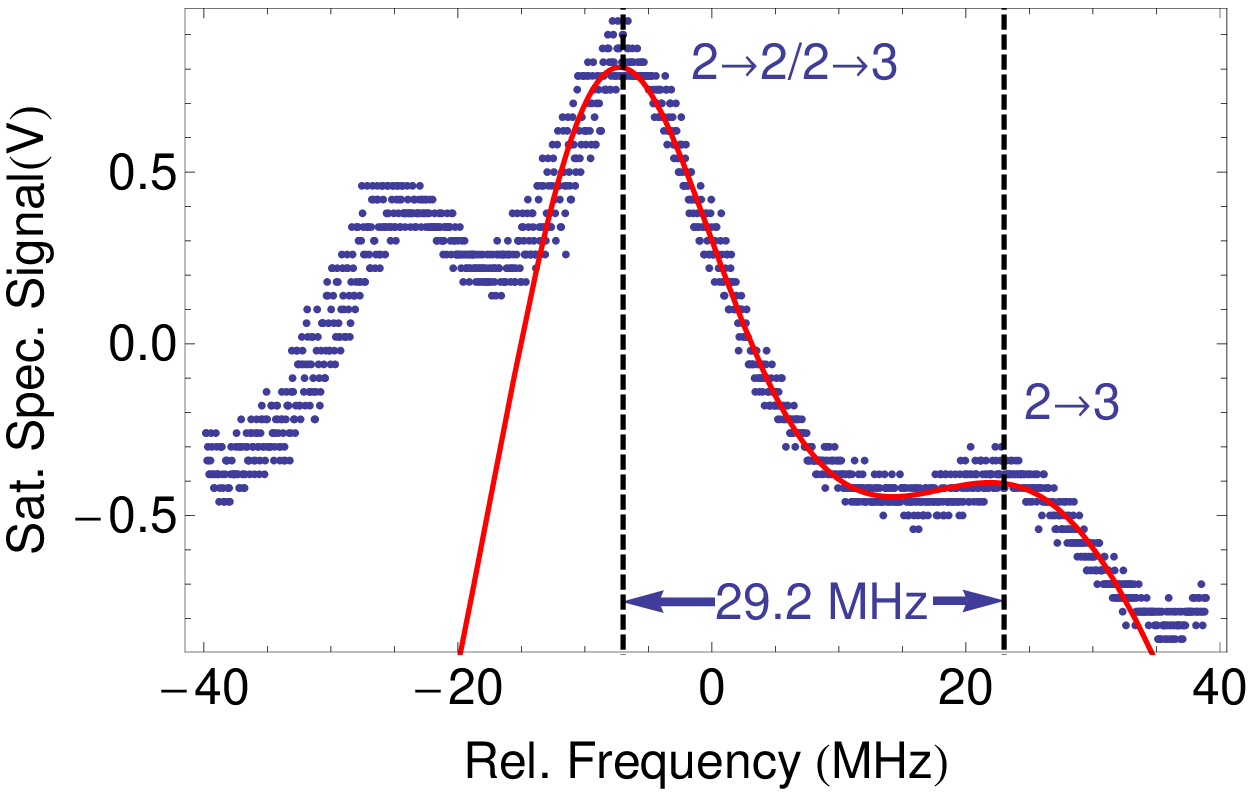}\\
	\includegraphics[width=0.9\linewidth]{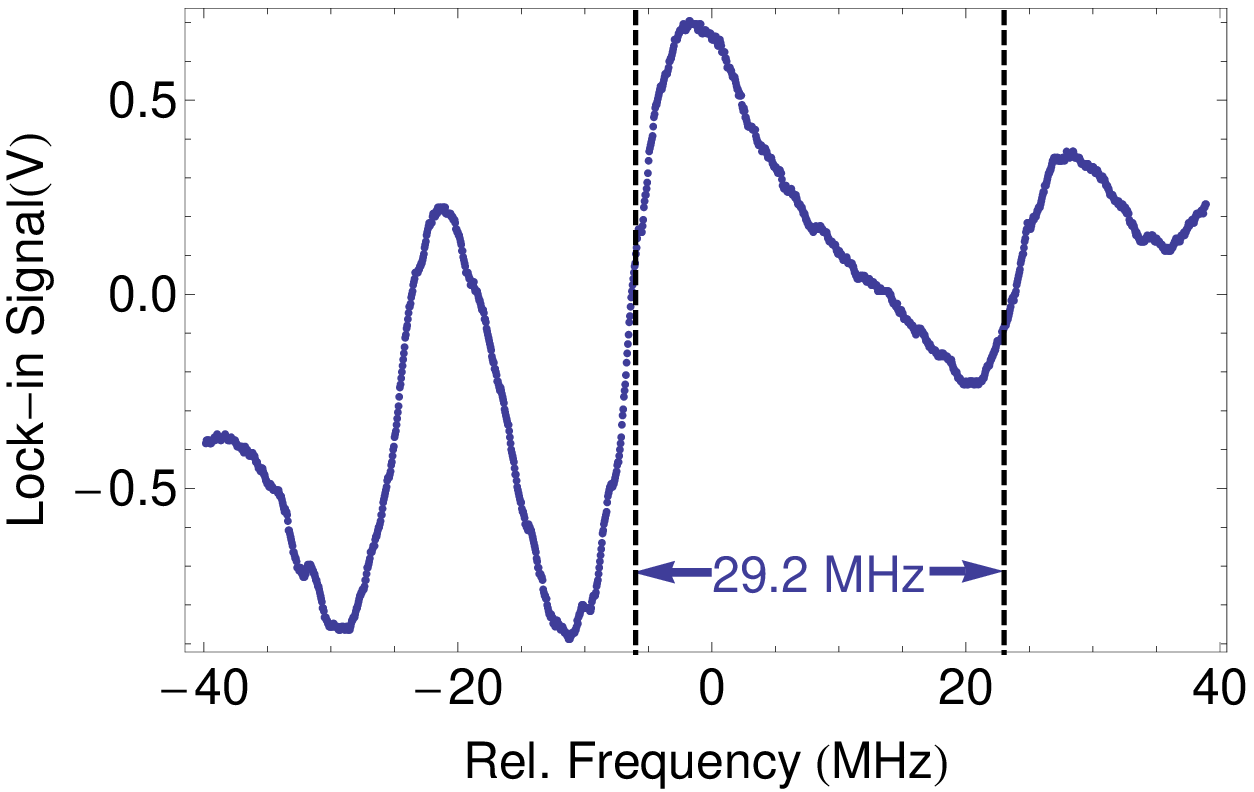}
	\caption{\label{fig:satspec} The saturation spectrum for stabilizing the 589 nm reference laser (top panel) has peaks at each of the hyperfine transitions and crossovers in sodium. Two known peaks were fit and the distance between them was calibrated to the known frequency separation of $29.163(43)\;\text{MHz}$.\cite{steck} The derivative (bottom panel) is used to lock to the peak of the central crossover resonance and the local slope at the lock point is used to calibrate the frequency jitter.
	}
	
\end{figure}

\subsection{Lock Stability}

\begin{figure}
	\includegraphics[width=0.9\linewidth]{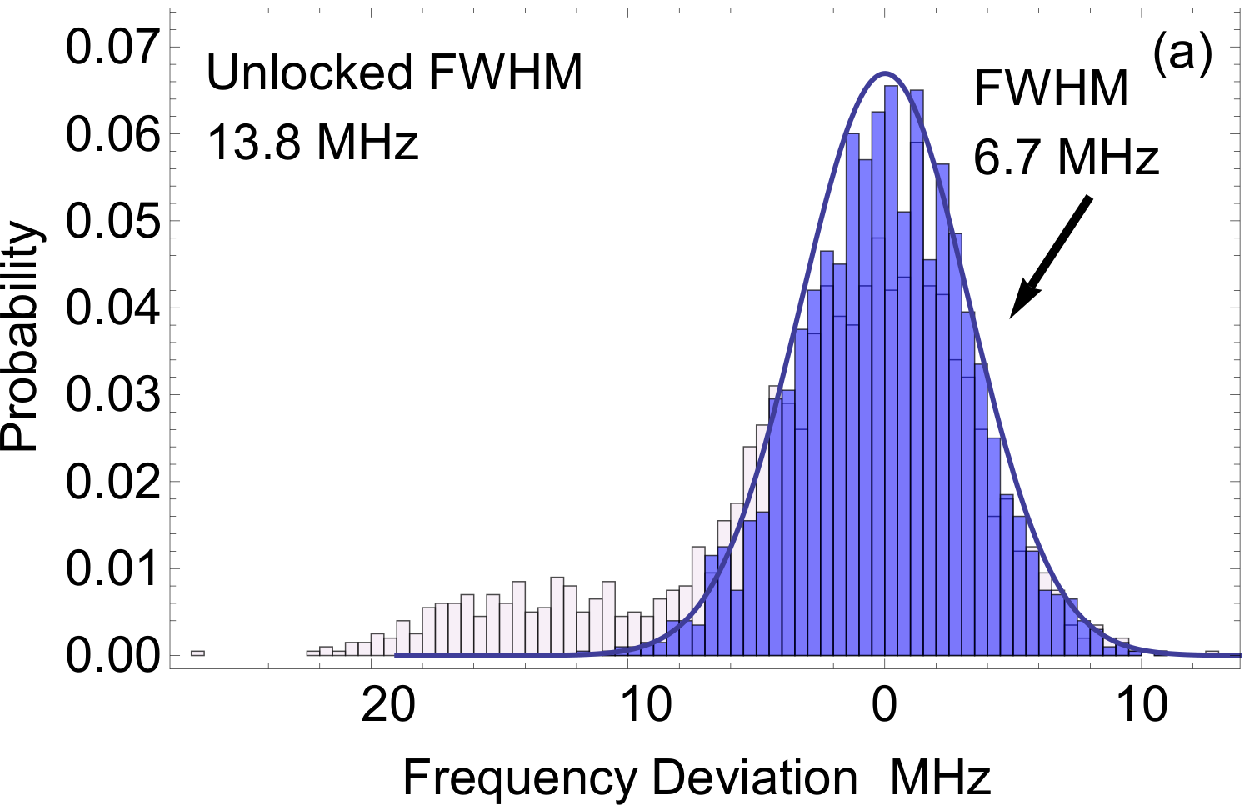}\\
	\vspace{5pt}
	\includegraphics[width=0.9\linewidth]{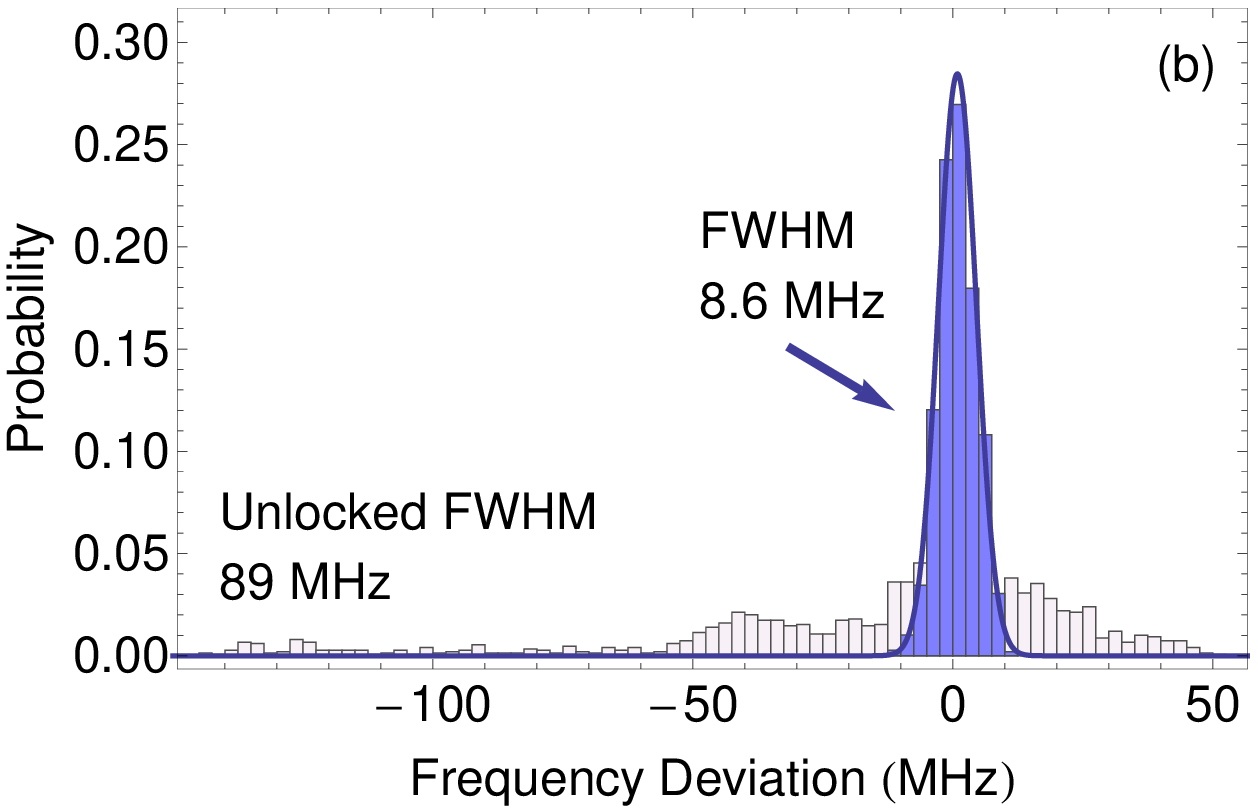}
	\caption{\label{fig:stab1} The 397 nm laser (a) and 866 nm laser (b) distributions are fit while locked and unlocked. The fit line indicates a fit of the locked laser frequency deviation to a normal distribution. Each laser was locked using a FP ramp frequency of 90 Hz, accumulating data for 1 hour in each case.}
\end{figure}

To accurately quantify the stability of the cooling lasers, we must first quantify the lock stability of the reference laser (589 nm). The reference laser is locked to the top of a peak in the Na Doppler-free saturation spectrum by using lock-in modulation and locking the derivative to the signal's zero crossing. Since we are locking to the top of a peak, there is a well-defined slope of the derivative signal at the locking point (0V, or slope 0), as seen in Fig.~\ref{fig:satspec}. We can use the known values of the transition hyperfine splitting between the $F=2\to 3'$ hyperfine transition and the crossover between the $F=2\to 2'\text{ and }F=2\to 3'$ transitions in the Na D2 line to calibrate how much our laser has deviated while locked. The difference between these two transition frequencies is half that of the difference between $F=2\to 2'$ and $F=2\to 3'$ transitions, which is known to be 58.326(43) MHz, therefore the distance between the transition and crossover as 29.163(22) MHz.\cite{steck} Using this as a reference and the local slope of the derivative at our lock point, a number entirely based upon our lock-in amplifier settings, we obtain the calibrated jitter of the locked laser. The distribution of the lock jitter, when sampled every 5 ms, while locked for 100 seconds, yields a FWHM linewidth of $\sim150$ kHz when the distribution was fit to a Gaussian profile, as seen in Fig.~\ref{fig:stab1}. The stability of the reference laser gives an upper-bound on the stability of the $\text{Ca}^+$ cooling diode lasers.

\begin{figure}
	\includegraphics[width=0.9\linewidth]{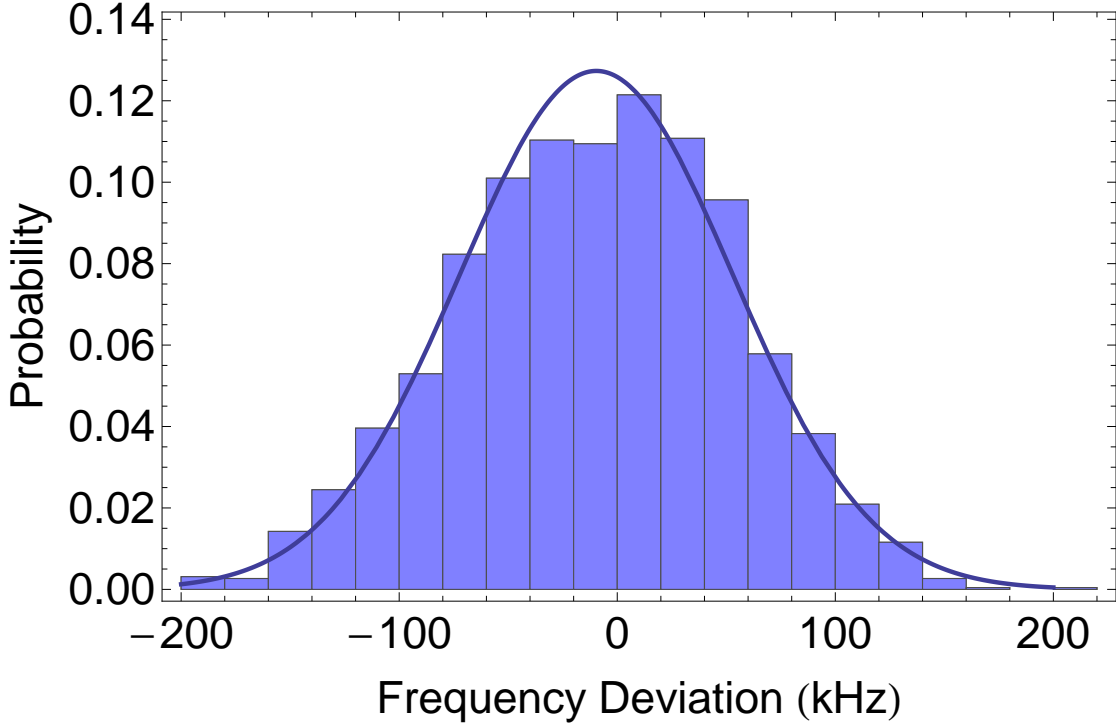}
	\caption{\label{fig:stab} The 589 nm reference laser has a long term linewidth of $\sim 150$ kHz as measured via the derivative lock signal. The bin size of the histogram is 20 kHz. These data were taken over the time period of 100 sec.}
\end{figure}

Our continuous scanning lock scheme requires enough long-term temperature stability that the features being scanned do not drift beyond the scan window, eventually wrapping around the scan. The Thorlabs FPs that we use exhibit thermal drift such that the cavity typically drifts less than a free spectral range in 3 hours. For certain applications that require a longer lock-time, a secondary lock could be implemented that fixes the position of the peak on the LabVIEW screen of the 589 nm reference beam relative to the FP ramp. This secondary lock could easily be implemented using a command in LabVIEW to adjust the ramp offset in order to fix the location of the 589 nm peak relative to the scan window. Unlike the transfer cavity lock or temperature stabilized FP cavities, this lock is internal to the LabVIEW program, and would not require any extra equipment.

The cooling and repumping transitions are at 396.847 nm and 866.216 nm in vacuum, with their respective transition partial widths of $2\pi\times 20.7 \text{ MHz}$ and $2\pi\times 1.7 \text{ MHz}$.\cite{caion} Each laser is stabilized to better precision than the transition linewidth in order to effectively laser-cool the ions. The stabilities of these two lasers were determined by compiling the locked and unlocked frequency data over the course of an hour. The FP signal jitter was measured and calibrated using the FP free spectral range. As we can see in Fig.~\ref{fig:stab1} (a), the 397 nm laser is very stable without a lock, resulting in the lock only slightly improving the laser linewidth from $2\pi\times13.8(2)$ MHz to $2\pi\times6.7(2)$ MHz, but much smaller than the transition linewidth of $2\pi\times20.7$ MHz. In Fig.~\ref{fig:stab1} (b), the custom-built 866 nm laser exhibits dramatic improvements when locked, reducing its linewidth from $2\pi\times89.1(5)$ MHz to $2\pi\times8.6(2)$ MHz, somewhat smaller than the total natural linewidth of $2\pi\times22.4$ MHz. This is more than adequate to repump the atoms into the cooling cycle. While our method does not lock well enough for applications requiring sub-MHz precision, it is   robust enough to lock the Doppler-cooling and repumping transitions in $\text{Ca}^+$ ions, and we have achieved smaller linewidths than similar techniques\cite{Nizamani:2013} previously reported.

\section{Conclusions}
We have successfully implemented a simple laser locking technique which simultaneously locks two lasers in widely different optical wavelength regimes to a reference laser. In principle, one could lock multiple lasers in the same way. We accomplish the locking by fixing the relative frequency difference between two lasers via their associated FP scan signals. The data acquisition and error signal generation used software generated via LabVIEW, using an internal digital PID locking module with the difference between laser frequencies measured in a FP cavity as the digital locking parameter. Our locking system, whose response time is limited by the speed at which the LabVIEW program iterates, yielded long-term stability of our commercial 397 nm laser of $2\pi\times6.7(2)$ MHz and the custom-built 866 nm laser to $2\pi\times8.6(2)$ MHz, both acceptable ranges compared to the natural linewidths for our experimental application of laser cooling $\text{Ca}^+$. 

\begin{acknowledgments}
Support in part from NSF grant PHY-1307874 is gratefully acknowledged.
\end{acknowledgments}

\end{document}